\documentclass{article}
\usepackage{spconf}
\usepackage{url}
\usepackage{graphicx}
\usepackage{amsmath}
\usepackage[mathscr]{euscript}
\usepackage{pifont}
\usepackage{wasysym}
\usepackage{amssymb}
\usepackage{xcolor}

\newcommand{\cmark}{\ding{51}}%
\newcommand{\xmark}{\ding{55}}%

\title{A Closer Look at Wav2Vec2 Embeddings for On-Device\\
Single-Channel Speech Enhancement}
\name{Ravi Shankar$^1$, Ke Tan$^2$, Buye Xu$^2$, Anurag Kumar$^2$}
\address{
  $^1$Department of Electrical and Computer Engineering, Johns Hopkins University, USA\\
  $^2$Meta Reality Labs, Redmond, Washington, USA}
\begin{document}

\maketitle
 
\begin{abstract}
Self-supervised learned models have been found to be very effective for certain speech tasks such as automatic speech recognition, speaker identification, keyword spotting and others. While the features are undeniably useful in speech recognition and associated tasks, their utility in speech enhancement systems is yet to be firmly established, and perhaps not properly understood. In this paper, we investigate the uses of SSL representations for single-channel speech enhancement in challenging conditions and find that they add very little value for the enhancement task. Our constraints are designed around on-device real-time speech enhancement -- model is causal, the compute footprint is small. Additionally, we focus on low SNR conditions where such models struggle to provide good enhancement. In order to systematically examine how SSL representations impact performance of such enhancement models, we propose a variety of techniques to utilize these embeddings which include different forms of knowledge-distillation and  pre-training. 

\end{abstract}
\noindent\textbf{Index Terms}: Speech Enhancement, Wav2Vec2, GCRN, Pre-training, Knowledge Distillation, Conditioning 

\section{Introduction}
Speech enhancement (SE) is a fundamental problems in the domain of speech signal processing. Broadly speaking, its goal is to enhance the quality (naturalness) and intelligibility of any given speech signal with or without making apriori assumptions about the noise or other distortions in the signal. SE systems have multiple applications such as noise suppression in phone calls, better communication in noisy environment, and in designing more robust hearing aids~\cite{enhancement_applications}. 

Speech enhancement is a very challenging problem to solve, mainly due to the blind nature of the noise (in the statistical sense), non-stationarity of the signal and, the duality in type of signal corruption, i.e., additive "vs" convolutional. Having said that, significant improvements have been made in recent times in separating noise-like component from speech using supervised machine learning methods. Common techniques for speech enhancement formulates it as a discriminative task where the goal might be to learn a mask or directly predict the clean speech~\cite{cIRM}. This can be done in time-domain as well as time-frequency (TF) domain.

To this end, multiple novel neural network architectures have been proposed such as~\cite{defossez2020real, hao2020fullsubnet, le2021dpcrn, DCN_main, tan2019learning, DCNUNET, DCCRN, demucs, fullsubnet}. Generative modeling via either score matching or denoising diffusion methods have also been proposed to synthesize clean speech from noisy inputs \cite{diffusion_enhancement, diffusion_enhancement_2, generative_enhancement, conditional_enhancement}. Beyond supervised training of deep neural networks, some recent works have explored semi and self-supervised approaches~\cite{triplet_enhancement, tzinis2022remixit, xiang2020parallel, fujimura2021noisy, cheng2021improving, zezario2020self}.

Among the model architectures proposed in the above works, gated convolutional recurrent network (GCRN) proposed in~\cite{tan2019learning}, is a special type of hybrid convolutional-recurrent model for enhancement in time-frequency domain. This model consists of a gated convolutions~\cite{gated_conv} followed by grouped long short-term memory (LSTM) layers to reduce parameters in a U-net~\cite{unet} style architecture. The architecture can be conveniently used to formalize speech enhancement using various targets (mask, magnitude spectrogram, complex spectrogram) and loss functions to obtain state-of-the-art speech enhancement system. Moreover, the encoder-decoder design of GCRN makes it suitable for our study. Hence, we use a GCRN architecture based model in this study.

\begin{figure*}[!t]
\centering
\includegraphics[width=0.89\textwidth, height=5.4cm]{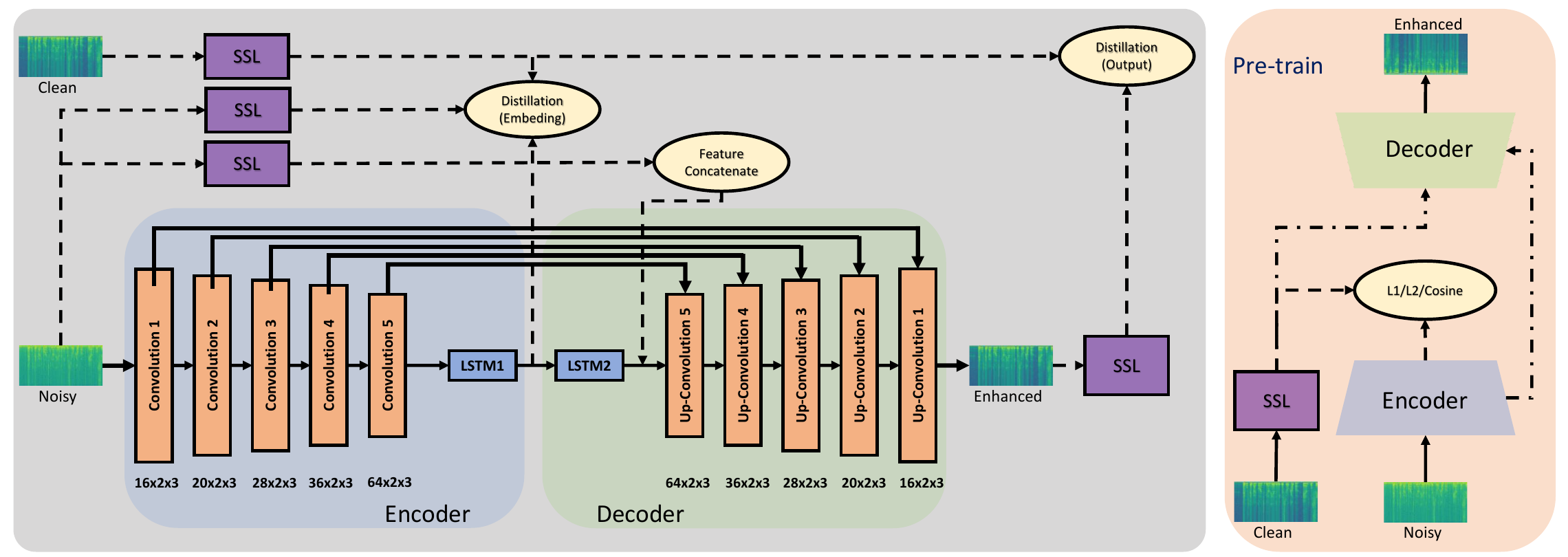}
\caption{GCRN model and different modes of using SSL embeddings to guide enhancement model. The right panel (red) shows our pre-training modes. The left panel (green) shows knowledge distillation modes and uses of SSL embeddings as inputs.}
\label{fig:GCRN_model}
\vspace{-3mm}
\end{figure*}

Recent years have also seen development of a variety of works on speech representation learning and it's application to different tasks. The goal of representation learning is to extract meaningful features from a signal (audio/video/text) in a self-supervised/unsupervised manner. These learned representations are helpful for downstream tasks. Some of the most popular models for speech representation learning are Wav2Vec2~\cite{baevski2020wav2vec}, HuBERT~\cite{hsu2021hubert} and WavLM~\cite{chen2021wavlm}. With similar underlying neural architectures, they are trained in slightly different ways. The key objective of all these models is to capture the broad phonetic-linguistic structure in the speech.  

While SSL models have been found useful in ASR tasks, only a few works, ~\cite{boosting, investigate_SSL, yossi}  have focused on explicitly using self-supervised learned features for speech enhancement. One rationale behind using SSL representations for enhancement can be to inject phonetic information which have been found to be useful for enhancement ~\cite{phonetic_enhancement}. \cite{boosting, investigate_SSL, yossi} use these SSL embeddings as inputs to train the enhancement model, either alone or in concatenation with short-time Fourier transform (STFT) representations. The authors in~\cite{yossi} use SSL embeddings to supervise and regularize their enhancement network. In above works, the improvements from SSL models are fairly limited, and they fail to provide clear explanation for their results. 

Our goal in this paper is to systematically investigate different ways of using SSL embeddings to improve an SE system. More specifically, we focus on on-device and real-time processing which constrains how SSL embeddings can be used. Such SE systems are expected to be \emph{(a)} causal - no future look ahead, and, \emph{(b)} of low compute footprint. The GCRN neural architecture can be used to design and develop an SE models satisfying these characteristics. However, they may suffer from unsatisfactory performances in low-SNR conditions~\cite{on_device_lowsnr}.  Hence, the key question we study is - \textbf{Can SSL embeddings improve on-device SE systems in low-SNR conditions?}. In particular, we study the popular wav2vec2.0 SSL model and attempt to utilize it to improve a GCRN based on-device SE model. 

%We lay down the following operating conditions for this study: \emph{(a)} causal model for enhancement, \emph{(b)} low compute footprint models for on-device support and, \emph{(c)} low signal to noise ratio (SNR) (e.g -5db) for adverse acoustic environments. This begs the question: \textbf{why study enhancement under such strict setting?} Current neural network based enhancement models perform well at relatively high SNRs. Moreover, they are mostly non-causal in nature with millions of parameters, which makes them inefficient for operations where the compute budget is severely limited. Hence, it is imperative to understand how to use Wav2Vec2 embeddings for speech enhancement under these challenging scenarios. 

The conditions outlined above constrain how any SSL model can be used for enhancement. SSL models are usually very large, non-causal and hence fine-tuning them~\cite{fine_tune_hubert} is not a possible path for using them in our case.  More generally, feeding SSL embeddings as inputs (or any form of conditioning) to the enhancement model will break causality as well as compute constraints as the SSL model will be needed during inference.  %Similarly, even the base Wav2Vec2 model is fairly large ($\sim$90M parameters), which means that feeding embeddings as inputs to the enhancement model will dramatically increases the computational cost during inference. 
Hence, the core idea which motivates the solution space here is that \emph{any approach to use SSL models for enhancement should not change the behavior of the underlying SE model during inference}, that is if the original SE model is causal and small, uses of SSL models to improve them should not change these characteristics. 

Keeping the above motivation in mind, we propose and analyze different ways in which an SSL model can be used to improve an enhancement model. Our approaches include those based on using SSL models as teachers for knowledge distillation as well as for pre-training of enhancement models. Along with comprehensive quantative analysis we also try to bring an understanding of Wav2vec2 structure.  We show that it is difficult to transfer the structure and information captured by Wav2Vec2 model to small enhancement models.  
\begin{figure*}[!t]
\centering
\includegraphics[width=0.70\textwidth, height=4cm]{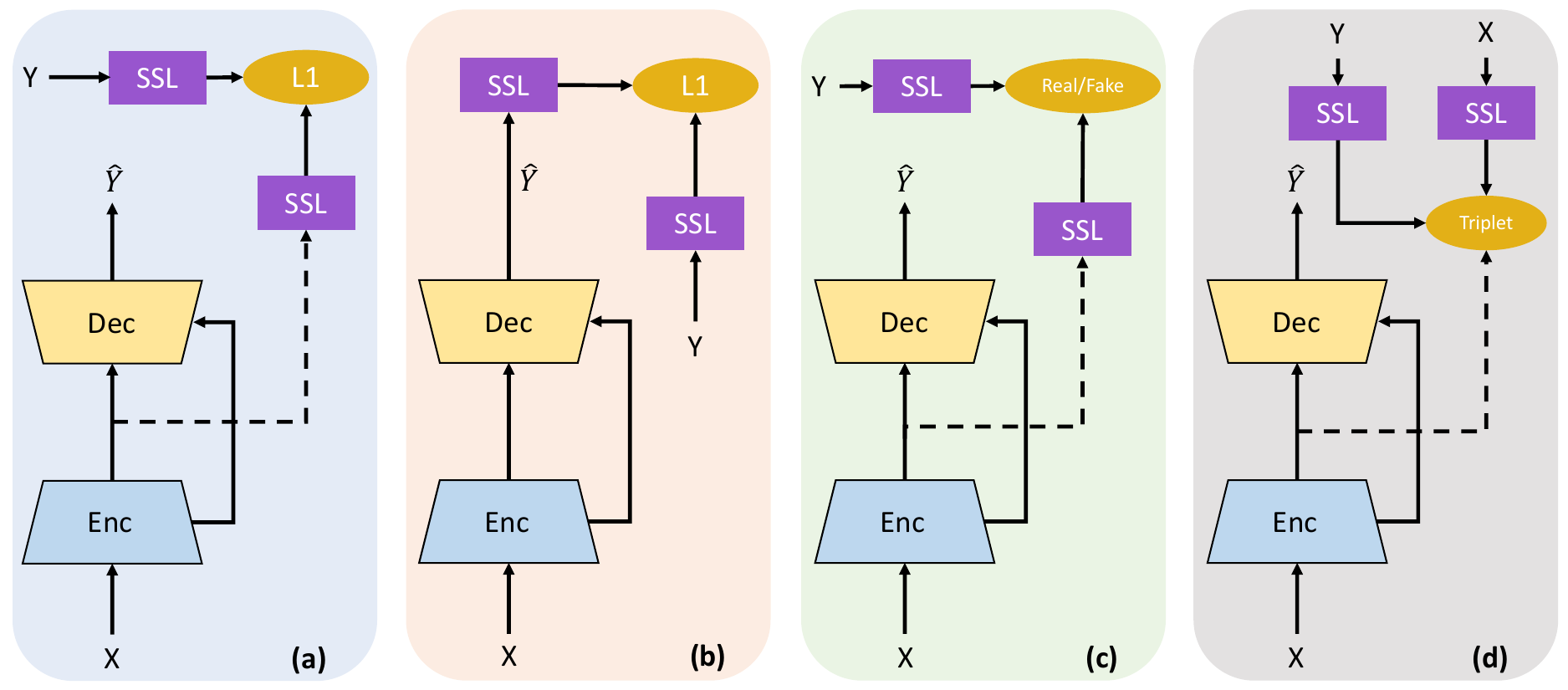}
\caption{Different ways of knowledge distillation from Wav2Vec2 embeddings used in this paper. (a) Sample-wise distillation via encoder output, (b) distillation via enhanced signal (c) adversarial distillation, and (d) triplet loss based distillation.}
\label{fig:distil_modes}
\vspace{-2mm}
\end{figure*}

%\vspace{-12mm}
\section{Method}
\vspace{-2mm}
In this section, we describe different approaches for using SSL model for enhancement. The input to each of these models is the spectrogram representation of speech signal extracted using a window of length $25$ms with a $20$ms stride to achieve the downsampling factor of 320. Note that, this choice is made for feature extraction to be consistent with the Wav2Vec2 model. Note: \emph{In our experiments, we use the Wav2vec2 model as SSL model. Hence, SSL and Wav2Vec2 embeddings are used interchangeably in this paper}. 

\vspace{-4mm}
\subsection{Overview}
\vspace{-2mm}
An overview of our framework is shown in Fig. \ref{fig:GCRN_model}. Our base enhancement model consists of a GCRN (details in Sec \ref{sec:base_se}) model. We propose 3 approaches to employ the SSL model to improve our enhancement model. (a) \emph{Feature Concatenate: } (Sec. \ref{sec:se_kd}) In this case the SSL embeddings are used to condition the decoder. Clearly, providing SSL embeddings as input to the GCRN model would make the overall inference non-causal and of large compute, breaking the constraint outlined earlier. Since this method does not satisfy the requirements it is a baseline for comparison. (b) \emph{Knowledge Distillation: } (Sec. \ref{sec:se_kd}/\ref{sec:kd_se_out}) Employing a teacher-student framework we propose a variety of ways to distill knowledge from the SSL model to the enhancement model. (c) \emph{Pre-training: } Lastly, we use the SSL model to pre-train the enhancement model. 

\vspace{-4mm}
\subsection{Baseline Enhancement Model}
\label{sec:base_se}
\vspace{-2mm}
The baseline enhancement framework used in this paper is a causal GCRN model which learns a complex spectral mapping from noisy speech to clean speech~\cite{tan2019learning}. The GCRN~(Fig.\ref{fig:GCRN_model}) consists of a stack of down-sampling convolutional layers followed by 2 layers of uni-directional LSTMs. The output of these LSTMs are then up-sampled by a set of transposed convolutions and added to the residue from down-sampling layers for generating final output. The causal structure of GCRN facilitates streaming capability for continuous on-the-fly operation. Furthermore, the recurrent operation in GCRN is performed group-wise (along feature dimension) to reduce the number of trainable parameters to $<\;$4M resulting in a memory footprint of only 16 megabytes. 

Mathematically, we denote the noisy speech as $\mathbf{X}$, clean speech as $\mathbf{Y}$ and the enhancement model as $f_e$. The training objective is to maximize $\mathscr{L} = SISDR(f_e(\mathbf{X}), \mathbf{Y})$ with respect to parameters of $f_e$. Scale-invariant signal-to-distortion ratio (SISDR)~\cite{sisdr} is defined via the following equation: 
\vspace{-2mm}
\begin{equation}
    \small{\text{$ SISDR(f_e(\mathbf{X}), \mathbf{Y}) = 10\log_{10}\frac{\Arrowvert\alpha \mathbf{Y}\Arrowvert^2}{\Arrowvert \alpha \mathbf{Y} - f_e(\mathbf{X}) \Arrowvert^2},$}}
    \vspace{-2mm}
\end{equation}
where $\alpha = \frac{f_e(\mathbf{X})^T \mathbf{Y}}{\Arrowvert \mathbf{Y} \Arrowvert^2}$.

Left panel in Fig. \ref{fig:GCRN_model} shows the three approaches by which SSL embeddings are used to guide the SE model. The details are described in Sec. \ref{sec:se_in} and Sec. \ref{sec:se_kd}.

\vspace{-4mm}
\subsection{Wav2Vec-2 Embeddings}
\vspace{-2mm}
Wav2Vec2~\cite{baevski2020wav2vec} is one of the most popular self-supervised learned speech model. The model consists of a stack of convolutional layers to exploit short-time stationarity of speech followed by a stack of multi-head attention layers. A vector quantization layer discretize the convolutional feature space. By masking the output representation at random time-frames, the network is trained via a contrastive loss to maximize its similarity with the corresponding code-book vector from quantization layer. A maximum entropy criterion is added to the loss function to encourage equal uses of each code-book. 
%Therefore, a joint optimization over the discrete vectors and the representation makes it powerful for learning phonetic content of the signal. 

In this paper, we use Wav2Vec2 embeddings in two different ways: \emph{(a)} by using the last transformer layer output, and \emph{(b)} by convex combination of multi-layered outputs where weights are estimated ad-hoc. The rationale behind using multiple layer outputs hinges on the observations made by~\cite{ssl_tts, neural_analysis} that, intermediate features can store important para-linguistic information for speech reconstruction.

\vspace{-4mm}
\subsection{Wav2Vec2 Embeddings as Input}
\label{sec:se_in}
\vspace{-2mm}
The simplest approach for using SSL embeddings is providing them as an extra input to the enhancement model. This conditioning is done via concatenating Wav2Vec2 features with noisy speech spectrogram which are then fed to the SE model. Prior works~\cite{boosting} have shown that concatenation of SSL after the bottleneck LSTM layers work better as it provides useful phonetic guidance right before the generative up-sampling operation begins. Hence, we adopt the same strategy in our experiments. The concatenated features (bottleneck features + SSL) are passed via a linear projection layer to maintain the dimensionality for residual skip connections. We further investigate two different strategies for concatenating the embeddings: (a) using the SSL embedding from the last layer of Wav2Vec2 model and (b) using weighted combination of embeddings obtained from each self-attention layer calculated adaptively. Let $g_s$ be the SSL model then, we minimize $\mathscr{L}_E = -SISDR(f_e[\mathbf{X}, g_s(\mathbf{X})], \mathbf{Y})$, where $g_s(\mathbf{X})$ is the feature extracted from Wav2Vec2 model.

We refer to this approach as \emph{feature concatenation} since the SSL embeddings are concatenated with LSTM outputs. Note that, in this case, \emph{Wav2Vec2 embeddings would be required during inference and making the overall system large and non-causal}. 
\vspace{-4mm}
\subsection{Distillation to SE Embeddings}
\label{sec:se_kd}
\subsubsection{Distillation via L1 Loss}
In this approach, the Wav2Vec2 representations of the target clean speech are directly used to inject knowledge in the SE model. It forces the outputs of the SE encoders to have the same semantic information as SSL embeddings of the clean speech (Fig.~\ref{fig:distil_modes}(a)). %The underlying hypothesis here is that the output of LSTM layers in SE models should retain language-specific knowledge for reconstruction by up-convolution. 
Note that, this technique uses the Wav2Vec2 embeddings extracted from the ground-truth signal itself. Therefore, it can act as a stronger prior than the naive concatenation because self-supervised models use clean speech to learn their parameters. The outputs of the LSTM layers which are decoded to produce the clean speech are expected to be enriched through the SSL representations. 

We employ a linear projection layer to match dimensionality of bottleneck features with the SSL embeddings which are learned via backpropagation. Similar to the concatenate mode, we use two forms of SSL embeddings, i.e., output of the last layer, and weighted combination of each layer of Wav2Vec2. The overall loss function for training in this case is:
\vspace{-1mm}
\begin{equation}
    \small{\text{$\mathscr{L}_E = -SISDR(f_e(\mathbf{X}), \mathbf{Y}) \; + \; \lambda \times \Arrowvert g_s(\mathbf{Y}) - f_e^{enc}(\mathbf{X}) \Arrowvert$}}
\end{equation}

\subsubsection{Distillation via Adversarial Loss}
The previous distillation approach use sample-wise similarity constraint that puts a strong prior on the enhancement network. We propose another way of doing knowledge distillation via distribution matching enforced through objectives such as KL-divergence penalty. We experiment with the distribution level matching of the Wav2Vec2 embeddings with GCRN encoder via an adversarial loss term (Fig.~\ref{fig:distil_modes}(c)). We use a block-wise convolutional discriminator proposed in~\cite{hifi_gan} with a period of 1 to consider each frame of the embeddings. Mathematically, denoting the discriminator network by $\mathscr{D}$, the enhancement and discriminator objective are written as:
\begin{align}
    \small{\text{$\mathscr{L}_E = -SISDR(f_e(\mathbf{X}), \mathbf{Y}) \; + \; \lambda \times \Arrowvert \mathscr{D}(f_e^{enc}(\mathbf{X})) - 1 \Arrowvert_2^2$}} \nonumber \\
    \text{and, } \; \small{\text{$\mathscr{L}_D = \frac{1}{2}\Arrowvert \mathscr{D}(f_e^{enc}(\mathbf{X})) - 0 \Arrowvert_2^2 \; + \; \frac{1}{2}\Arrowvert \mathscr{D}(g_s(\mathbf{Y}) - 1 \Arrowvert_2^2$}}
\end{align}
This is the least-square formulation of GAN~\cite{lsgan} which has been shown to minimize the Pearson $\chi^2$ divergence between the Wav2Vec2 embeddings and GCRN encoder output. 

\subsubsection{Distillation via Triplet Loss}
Triplet loss is yet another way to enforce a similarity between the GCRN encoder embeddings and Wav2Vec2 representations in the manifold space. Authors in~\cite{triplet_enhancement} proposed triplet loss in the context of unsupervised training of speech enhancement. The underlying idea in triplet loss is to maximize the margin (up to a certain threshold) between the set of GCRN embeddings and embeddings obtained for clean and noisy speech from Wav2Vec2 (Fig.~\ref{fig:distil_modes}(d)). Note that, this is a stronger penalty than contrastive loss which only forces dissimilarity among different representations. The objective is:   
\begin{align}
    \small{\text{$\mathscr{L}_E = -SISDR(f_e(\mathbf{X}), \mathbf{Y}) \; + \lambda \times \mathscr{L}_T \qquad \qquad$}} \nonumber \\
    \small{\text{where, $\mathscr{L}_T = \max(\Arrowvert \mathbf{a} - \mathbf{p} \Arrowvert - \Arrowvert \mathbf{a} - \mathbf{n} \Arrowvert + m, 0) \qquad \qquad$}}
\end{align}
Here, $\mathbf{a} = f_e^{enc}(\mathbf{X})$ represents the GCRN encoder representations (also called anchor), $\mathbf{p} = g_s(\mathbf{Y})$ is the Wav2Vec2 embeddings from clean speech and $\mathbf{n} = g_s(\mathbf{X})$ is the same from noisy speech. The margin $m$ is set to $100$ in this task to account for high dimensionality of the embeddings.
\vspace{-3mm}
\subsection{Distillation to SE Outputs}
\label{sec:kd_se_out}
Another distillation approach is to enforce similarity in the latent space of the enhanced and the clean speech. We do this by adding an extra loss term to the objective function which forces similarity between Wav2Vec2 representations of the enhanced signal and the ground-truth speech (Fig.~\ref{fig:distil_modes}(b)). The underlying hypothesis is same as the previous method, i.e., the enhanced speech should have the same phonetic-linguistic content as the clean speech, thereby, improving its intelligibility. This method does not require additional linear layer, but it does require gradient backpropagation through the SSL model during training. The overall loss function is given by: 
\vspace{-1mm}
\begin{equation}
    \small{\text{$\mathscr{L}_E = -SISDR(f_e(\mathbf{X}), \mathbf{Y}) \; + \; \lambda \times \Arrowvert g_s(\mathbf{Y}) - g_s(f_e(\mathbf{X})) \Arrowvert$}}
\end{equation}

\subsection{Pre-training via Wav2Vec2 Embeddings}
SSL is primarily used as unsupervised pre-training strategy, models are then fine-tuned on different downstream tasks. However, our goal here is to use the Wav2Vec2 model to improve the given SE network under consideration. To achieve this, we propose to pre-train the GCRN enhancement model using SSL embeddings. The general motivation is that pre-training provide better initialization of the model parameters than random. This approach is illustrated in right panel of Fig. \ref{fig:GCRN_model}. GCRN facilitates a straightforward decomposition of the model architecture into an encoder and a decoder. The LSTM blocks in the bottleneck layers are split equally among the encoder and decoder components. We pre-train both the encoder and the decoder. 

\textbf{Encoder Pre-training:} For pre-training of the encoder, we provide noisy spectrogram as input and predict the Wav2Vec2 embeddings as output. This corresponds to knowledge distillation from the large scale SSL model to the encoder of the SE network.

\textbf{Decoder Pre-training:} We pre-train the decoder by formulating the task as a speech generation problem. To achieve this, the decoder is trained by making it predict the clean spectrogram conditioned on the ground truth SSL embeddings. We also experiment with the decoder training conditioned on the encoder outputs rather than ground truth embeddings. In the former case, the residual connection based on up-sampling operation is replaced by locally duplicating the learned features by a factor of 2. 

\subsection{Training Details}
We train the enhancement model with complex STFT features extracted from the noisy and clean speech pair. We use the Adam optimizer with a fixed learning rate of 0.001 for 4 million steps with a batch size of 200. SI-SDR loss between the ground truth and predicted signal is used for training. 

\section{Experiments and Results}
%We start this section with the description of the dataset used for running experiments followed by their respective results. 

\subsection{Dataset}
We use the DNS challenge~\cite{dns_noise} corpus in our experiments. The clean speech and the noise samples are mixed at random SNRs between -5dB and 5dB for training and validation set. The testing set consists of 500 samples of clean speech from Librispeech test-clean mixed with noise (from the test sets) at a fixed -5dB SNR to simulate challenging real-world scenario. 

\subsection{Baseline, Feature Concatenation and Distillation}
We first compare different approaches outlined in Section 2. Table.~\ref{tab:baseline} summarizes the result of this experiment. We can see that weighted sum concatenation (*-ws)  and distillation via output works best on all three metrics, i.e. PESQ, STOI and SI-SDR. The differences however, are very small in practice. The last column of Table~\ref{tab:baseline} shows whether the model complies with the constraint set in the beginning of this paper namely, causality and memory footprint. 
%The base model (without any SSL input) performs relatively well and is on par with the best model. 
As expected, distillation via adversarial loss performs poorly in comparison to other modes because distribution level prior is weak. Triplet loss however, performs similar to sample-wise modes since it enforces similar constraint as the SE embedding distillation with a margin component.  
\begin{table}[h]
\addtolength{\tabcolsep}{-3.5pt}
\begin{center}
\begin{tabular}{ |c c c c c| }
 \hline
 \textbf{\small{Model}} & \textbf{\small{PESQ}} & \textbf{\small{STOI}} & \textbf{\small{SI-SDR}} & \textbf{\small{Constr.}}\\ 
 \hline
 Base & $1.59$ & $0.84$ & $9.1$ & \cmark\\ 
 Feature Concat & $1.55$ & $0.83$ & $8.9$ & \xmark\\ 
 Feature Concat-ws & $1.60$ & $0.84$ & $9.3$ & \xmark \\ 
 Distillation Embed. & $1.52$ & $0.83$ & $9.1$ & \cmark \\ 
 Distillation Embed.-ws & $1.56$ & $0.83$ & $9.2$ & \cmark \\ 
 Distillation Output & $1.60$ & $0.84$ & $9.3$ & \cmark \\ 
 Distillation Adversarial & $1.52$ & $0.82$ & $7.5$ & \cmark \\
 Distillation Adversarial-ws & $1.55$ & $0.81$ & $8.4$ & \cmark \\
 Distillation Triplet & $1.56$ & $0.81$ & $8.5$ & \cmark \\
 Distillation Triplet-ws & $1.57$ & $0.83$ & $8.9$ & \cmark \\
 \hline
\end{tabular}
\end{center}
\caption{Baseline GCRN model and different techniques considered towards using Wav2Vec2 embeddings for enhancement. \textbf{Noisy PESQ: 1.11, STOI: 0.69, SI-SDR: -4.99dB}}
\vspace{-0mm}
\label{tab:baseline}
\end{table}

Overall, we can conclude that the Wav2Vec2 embeddings do not provide any significant improvement over the vanilla model under on-device requirement scenario. We hypothesize the main reasons why it fails to do so as follows: first, concatenation adds an overhead of extra parameters in the model which may change the loss landscape of the parameterized function making it harder to find any better local optima. Second, distillation via embedding act by adding a penalty on the latent space representation of noisy speech. This might be a weak signal because the Wav2Vec2 embeddings retain the qualitative aspects of speech in trace amounts~\cite{ssl_tts}. 
\vspace{-3mm}
\subsection{Pre-training with SSL}
We use the Wav2Vec2 embedding extracted from clean speech to train the GCRN encoder. The goal is to learn the latent representation of clean speech. We also pre-train the decoder from (a) encoder's output and (b) SSL embeddings directly to generate clean speech. In the latter case, the up-sampling operation is performed locally instead of using skip connections from encoder. We also experiment with different types of encoder losses namely, L1, L2 and Cosine metric and pick only the best one for final fine-tuning. The selection is based on which model performs best for enhancement. 
\begin{table}[h]
\addtolength{\tabcolsep}{-1pt}
\begin{center}
\begin{tabular}{ |c c c c| }
\hline
\textbf{\small{Encoder Loss/Decoder Input}} & \textbf{\small{PESQ}} & \textbf{\small{STOI}} & \textbf{\small{WER \%}}\\ 
\hline
L1/Frozen & $1.24$ & $0.74$ & $71.2$ \\
L1/Wav2Vec2 & $1.25$ & $0.83$ & $7.4$ \\
L2/Frozen & $1.22$ & $0.73$ & $76.7$ \\
L2/Wav2Vec2 & $1.24$ & $0.81$ & $16.5$ \\
Cosine/Frozen & $1.21$ & $0.72$ & $79.4$ \\
Cosine/Wav2Vec2 & $1.29$ & $0.82$ & $12.5$ \\
\hline
\end{tabular}
\end{center}
\caption{Pre-training of speech enhancement model using SSL. \textbf{Noisy PESQ: 1.11, STOI: 0.69, SI-SDR: -4.99dB}}
\vspace{-0mm}
\label{tab:pretraining}
\end{table}
\par
Table.~\ref{tab:pretraining} summarizes the result of enhancement task performed directly using the pre-trained models. Note that, we did not train this GCRN on any enhancement task - just encoder and decoder training as outlined before. Even though we do not train the model for noise removal, \emph{the model manages to do some form of enhancement}. This shows that a pre-training based on prediction of SSL representations can on it's own lead to some denoising capabilities. 

Further, we can see that when we train the decoder directly from SSL embeddings, the intelligibility of generated speech is higher. In fact, we obtain a word error rate of $<20\%$ upon decoding the reconstructed speech using a pre-trained CRDNN model from Speechbrain library~\cite{speechbrain}. Note that, the CRDNN model is trained on Librispeech 960-h corpus itself. Nevertheless, this shows that SSL captures the phonetic linguistic information but completely ignores other aspects of speech such as tonality, loudness and voice quality.

Finally, we pick the model trained with L1 loss on encoder for fine-tuning on the enhancement task. 
\begin{table}[h]
\addtolength{\tabcolsep}{-1pt}
\begin{center}
\begin{tabular}{ |c c c c| }
 \hline
 \textbf{\small{Encoder Loss / Decoder Input}} & \textbf{\small{PESQ}} & \textbf{\small{STOI}} & \textbf{\small{SI-SDR}} \\ 
 \hline
 L1/Frozen & $1.53$ & $0.83$ & $8.60$ \\ 
 L1/Wav2Vec2 & $1.54$ & $0.84$ & $8.90$ \\ 
 \hline
\end{tabular}
\end{center}
\caption{Speech enhancement assessment from fine-tuned models. \textbf{Noisy PESQ: 1.11, STOI: 0.69, SI-SDR: -4.99dB}}
\vspace{-0mm}
\label{tab:finetuning}
\end{table}
As we can see from (Table.~\ref{tab:finetuning}), pre-training provides no significant advantage over the base model for the speech enhancement task. In fact, the model performance worsens slightly compared to the GCRN model (see Table.~\ref{tab:baseline}) trained from scratch. We conjecture that this happens due to two main reasons: (a), the Wav2Vec2 embeddings only capture the information required for reconstruction of smoothed quantized features and need further fine-tuning for generative tasks such as enhancement. The evidence for this hypothesis is provided by Table.~\ref{tab:pretraining}. We can observe that the STOI scores are relatively high, meaning we get intelligible speech of poor quality when generated directly from Wav2Vec2 embeddings. This is in contrast to the enhancement task, where the representation learnt by encoder should encode both, the style and speaker information in conjunction with the phonetic/linguistic component of the speech utterance. Second, it is challenging to distill knowledge from large SSL models due to the inherent structure of embeddings themselves. We discuss this phenomenon in the next subsection. 
\begin{figure}[t]
\centerline{\includegraphics[width=0.95\columnwidth, height=3.1cm]{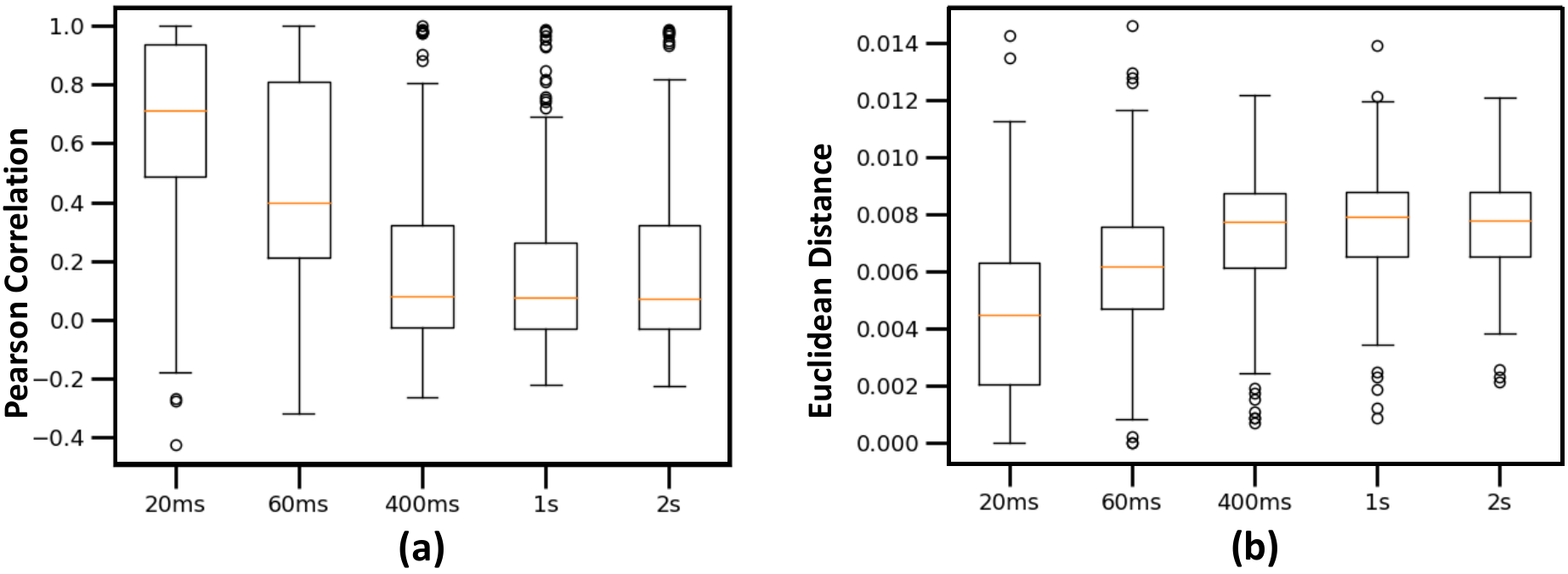}}
\vspace{-2mm}
\caption{\textbf{Wav2Vec2}: Box plot of (a) correlations and (b) Euclidean distances obtained from frames of Wav2Vec2 embeddings separated by 20ms, 60ms, 400ms, 1sec and 2sec.}
\label{fig:correlation_euclidean_SSL}
\vspace{-0mm}
\end{figure}
\vspace{-3mm}
\subsection{Structure of Wav2Vec2 embeddings}
We analyze the features from Wav2Vec2 for a variety of utterances and show the interesting correlation patterns in these embeddings. Fig.~\ref{fig:correlation_euclidean_SSL} shows the box plot of correlations and L2 norm between features separated by 20ms, 60ms, 400ms, 1sec and 2sec, respectively. We can see that the features are highly correlated up until 60ms (typical phoneme length), and are similar in magnitude (Euclidean distance) throughout an utterance leading to a conclusion that the phonetic/linguistic content is stored in the small magnitude variations between frames. The main implication of this pattern is that capturing such small differences is extremely difficult. Therefore, it forces the GCRN encoder to learn an average noise-like representation for each utterance in pre-training. 
\begin{figure}[t]
\centerline{\includegraphics[width=0.95\columnwidth, height=3.1cm]{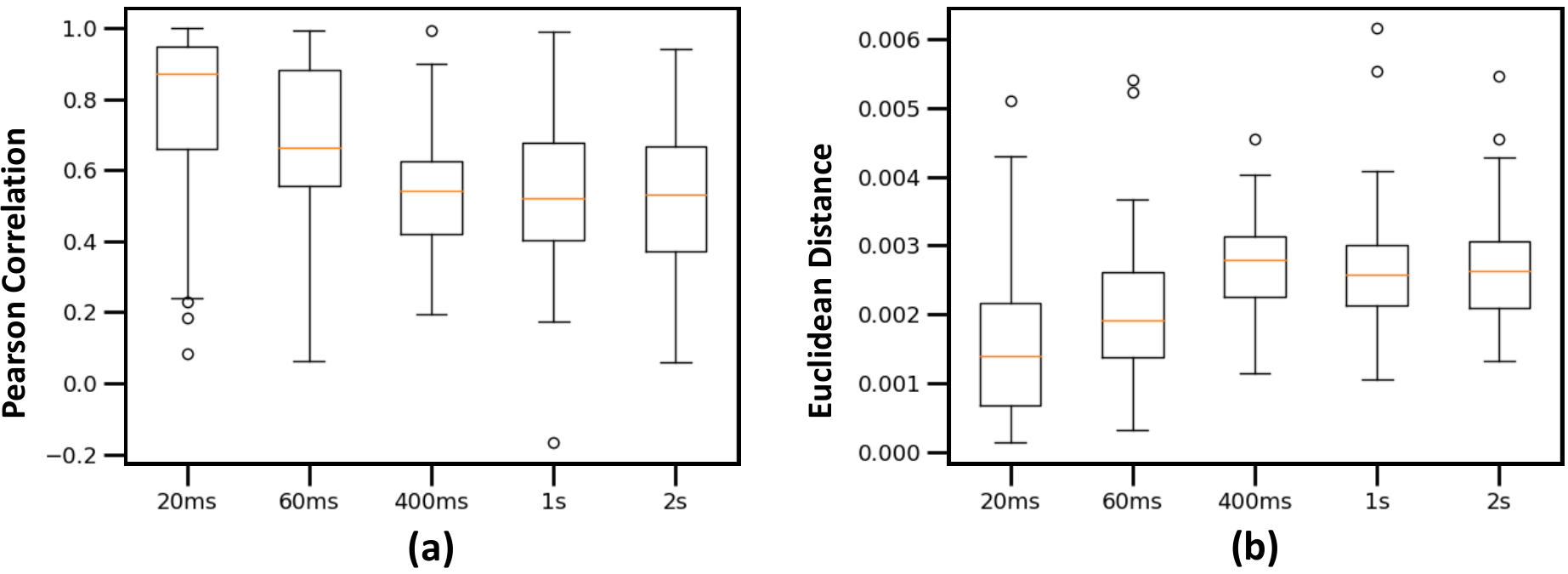}}
\vspace{-2mm}
\caption{\textbf{Distilled model}: Box plot of (a) correlations and (b) Euclidean distances obtained from frames of Wav2Vec2 embeddings separated by 20ms, 60ms, 400ms, 1sec and 2sec.}
\label{fig:correlation_euclidean_distilled}
\vspace{-3mm}
\end{figure}
\vspace{-3mm}
\subsection{Knowledge Distillation from SSL}
\begin{figure}[t]
\centerline{\includegraphics[width=\columnwidth, height=3cm]{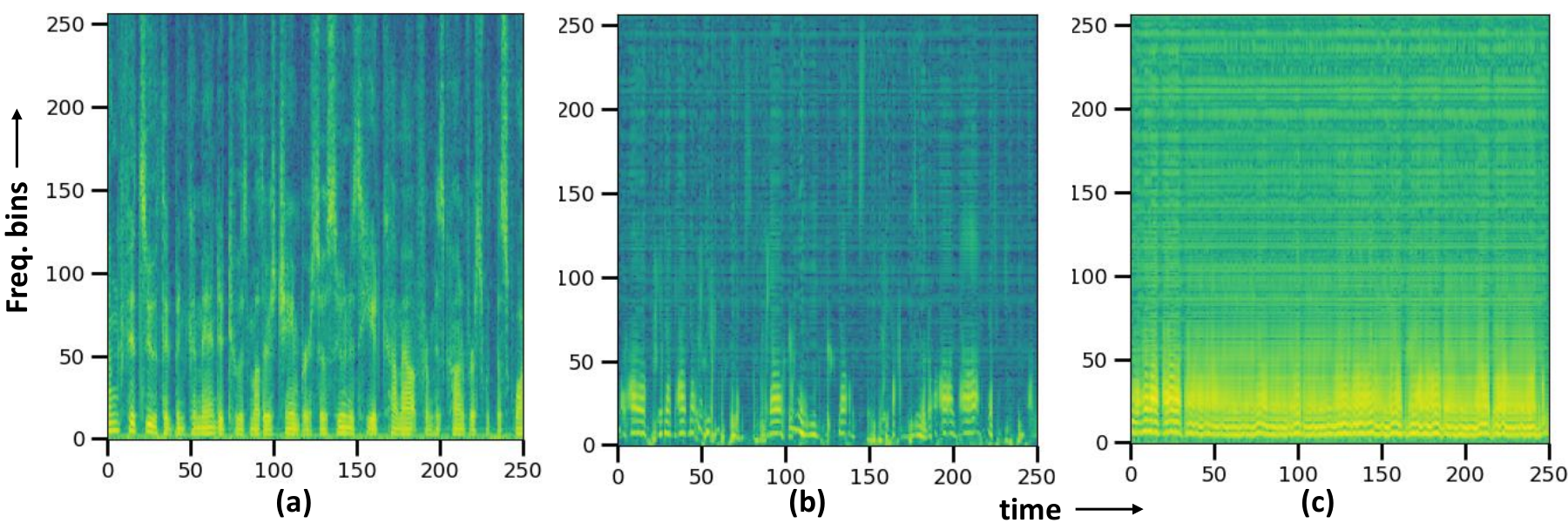}}
\vspace{-2mm}
\caption{Illustration of spectrograms: (a) ground truth speech (b) speech decoded using Wav2Vec2 embeddings and (c) speech decoded using embeddings extracted from the trained knowledge distillation model (same encoder as Wav2Vec2).}
\label{fig:decoded_SSL}
\vspace{-3mm}
\end{figure}

In this experiment, we probe into the question of finding out whether knowledge distillation from models such as Wav2Vec2 is possible or not. The GCRN encoder learns an average representation, but it could be due to the low complexity of the model itself. Therefore, we train a convolution-transformer stack identical to the Wav2Vec2 architecture itself, to predict the SSL embeddings from speech. We use a mix of L1 and cosine loss in this regard. Fig.~\ref{fig:correlation_euclidean_distilled} shows the correlation and Euclidean distance pattern of embeddings obtained from this new distilled model. Note that, it does not exhibit the same characteristics as the original embeddings from Fig.~\ref{fig:correlation_euclidean_SSL}. The change in correlation coefficient and normalized L2 distance is completely different from the pre-trained Wav2Vec2 model. Furthermore, Fig.~\ref{fig:decoded_SSL} shows an example generation from the GCRN decoder when prompted with original Wav2Vec2 embeddings (Fig.~\ref{fig:decoded_SSL}(b)) and the embeddings extracted from the distilled encoder (Fig.~\ref{fig:decoded_SSL}(c)). We can see that the original Wav2Vec2 embeddings allow the reconstruction of energy in the higher frequency bands whereas, the distilled model completely loses that information. The main reason for this behavior is the inability of even an expressive model to capture the intricate details present in the original embeddings. Therefore, we conclude that it is fairly challenging to distill knowledge from Wav2Vec2 model even if the student model follows the exact same architecture. 

\vspace{-4mm}
\section{Discussion and Conclusions}
\vspace{-2mm}
In this paper, we have explored different mechanisms to leverage Wav2Vec2 representation for the task of speech enhancement. We showed that under on-device constraints and low-SNR conditions, SSL models add little to  no value in improving the base enhancement model. We hypothesized that the SSL embeddings retain only the phonetic/linguistic component of speech and ignores the qualitative aspects of the signal. Our experiments with the pre-training of GCRN model using SSL embeddings further confirmed this hypothesis. We showed that GCRN decoder is capable of generating intelligible speech from Wav2Vec2 embeddings, but it lacks in quality as demonstrated by the poor PESQ and SI-SDR scores (Table~\ref{tab:pretraining}). In addition, we showed that the structure of these embeddings makes it difficult to pre-train the GCRN encoder. These features are difficult to reproduce even with a more expressive model, due to the phonetic details encoded in tiny variations across time. These subtle variations make it difficult for speech enhancement models to extract any meaningful information. Therefore, the Wav2Vec2 features require more refined understanding to use them for speech enhancement in challenging scenarios.    

\bibliographystyle{IEEEbib}
\bibliography{mybib}

\end{document}